\documentclass[twocolumn,showpacs,preprintnumbers,amsmath,amssymb,showkeys]{revtex4}
\usepackage{graphicx}% Include figure files
\usepackage{dcolumn}% Align table columns on decimal point
\usepackage{bm}% bold math
\newcommand{\bqa}{\begin{eqnarray}}
\newcommand{\eqa}{\end{eqnarray}}
\newcommand{\be}{\begin{equation}}
\newcommand{\ee}{\end{equation}}

\begin{document}

%\preprint{Preprint}

\title{$Z^+(4430)$ as a resonance in the $D_1(D_1^\prime)D^*$ channel}

\author{Ce Meng$~^{(a)}$ and Kuang-Ta Chao$~^{(a,b)}$}
\affiliation{ {\footnotesize (a)~Department of Physics, Peking
University,
 Beijing 100871, People's Republic of China}\\
{\footnotesize (b)~China Center of Advanced Science and Technology
(World Laboratory), Beijing 100080, People's Republic of China}}

%\date{\today}

\begin{abstract}
We study the possibility that the $Z^+(4430)$ state observed by
Belle is a $D_1D^*$ (or $D_1'D^*$) resonance in S-wave. Focusing on
its decays, We find that the open-charm decay into $D^*D^*\pi$ is
dominant. Furthermore, we use the re-scattering mechanism to study
its hidden-charm decays and find that the re-scattering effects are
significant in $D_1D^*$ channel but not in $D_1'D^*$ channel. For
the $J^P=1^-$ candidate, with chosen parameters, we can get
$\Gamma(Z^+\to\psi'\pi^+)/\Gamma(Z^+\to J/\psi\pi^+)\thickapprox
5.3$, which tends to account for why the $Z^+(4430)$ is difficult to
be found in $J/\psi\pi^+$. However, the $0^-$ candidate can not be
ruled out by our calculations.

% Valid PACS numbers may be entered using the \verb+\pacs{#1}+ command.
\end{abstract}

% PACS, the Physics and Astronomy
                             % Classification Scheme.

\maketitle

\section{Introduction}

Recently, there have been a number of exciting discoveries of new
hadron states in the hidden-charm sector. Among these new states,
the $Z^{\pm}(4430)$ may be the most special one, which was
discovered by the Belle collaboration~\cite{Belle07_Z4430} in the
invariant mass spectrum of $\psi'\pi^{\pm}$ in the decay $B \to
\psi'\pi^{\pm}K$ with mass $m_Z=4433\pm4\pm1$~MeV and width
$\Gamma_Z=44^{+17+30}_{-13-11}$~MeV.

The distinguishable feature of $Z^{\pm}(4430)$ is that it is a
charged resonance-like structure. Surely it can not be emplaced in
ordinary charmonium sector or explained by couple channel effects
between charmonium and open-charm thresholds. The closeness of $m_Z$
to the thresholds of $D^*(2010)D_1(2420)$ and $D^*(2010)D_1'(2430)$
may suggest that it is induced by the threshold effects of these
channels. However, it may not be a simple kinematic effect since it
is very narrow and the line-shape of $\psi'\pi$ distribution is very
sharp. Presumably, it is likely to be a resonance of $D^*D_1$ (or
$D^*D_1'$). Here, a "resonance" means a state formed in the
$D^*(2010)-D_1(2420)$ scattering by some inter-hadron forces (say,
meson exchanges), but not the bound state formed by the
colored-confining forces. Note that in the $B^+$ decays, the CKM
favored modes such as $B^+\to \bar{D^*}(2010)^0D_s^{*+}$ has a very
large branching ratio of about 2\%\cite{PDG06}. This can happen not
only for the $D_s^{*+}$ but also for the excited $D_s^{*+}$ system,
which can decay into $D^+K^0$ and $D^0K^+$ as well as
$D^+_1(2420)K^0$ and $D^0_1(2420)K^+$ with sizable branching ratios.
Then in the scattering process of $\bar{D^*}(2010)^0-D^+_1(2420)$
and $\bar{D^*}(2010)^0-D^0_1(2420)$, resonance-like structures $Z^+$
and $Z^0$ may be formed.

%One following question is whether or not it is bound. The answer is
%apt to NOT since it can not lies below the threshold safely due to
%the widths $\Gamma_{D_1}\sim20$ MeV  and
%$\Gamma_{D_1'}\sim300\mbox{-}500$ MeV are too large.
Here, we will not try to clarify the complicated dynamics in the
$D^*D_1^{(\prime)}$ scattering in this paper. However, it is worth
emphasizing that the configuration of $D^*D_1^{(\prime)}$ in $Z$
should favor S-wave because a centrifugal barrier will prevent the
formation of a resonance state. Then, the allowed $J^P$ of $Z$ is
$0^-$, $1^-$ or $2^-$. Thereinto, the possibility of $2^-$ may be
neglected since its production in $B\to Z(4430)K$ decay is
suppressed by the small phase space.

Another unusual feature about $Z(4430)$ is that there is no
report~\cite{Belle07_Z4430} on $Z^+\to J/\psi\pi^{+}$ in the
$J/\psi\to l^+ l^-$ mode, which implies the ratio
%&&&&&&&&&&&&&&&&&&&&&
\be\label{Ex:R_psi'/psi} R_{\psi'/\psi}=\frac{\Gamma(Z^+\to
\psi'\pi^{+})}{\Gamma(Z^+\to J/\psi\pi^{+})}>>1.
%\frac{{\mathcal
%B}(J/\psi\to l^+ l^-)}{{\mathcal B}(\psi'\to l^+ l^-)}\thickapprox8.
\ee
%&&&&&&&&&&&&&&&&&&&&&
It is not easy to understand this "$J/\psi$-suppressed problem"
since $J/\psi$ has many properties similar to $\psi'$ and the phase
space should favor the $J/\psi$ production.

In this paper, we will stress on the decays of $Z$ assuming that it
is a $D^*D_1^{(\prime)}$ resonance with $J^P=0^-$ or $1^-$. Since
$D_1^{(\prime)}$ decays to $D^*\pi$ dominantly, one can expect the
width of $Z$ is dominated by $\Gamma(Z\to D^*D^*\pi)$. On the other
hand, since $Z$ is discovered in $\psi'\pi$ system with a quite
large product branching ratio:

%&&&&&&&&&&&&&&&&&&&&&
\be\label{Ex:Br*Br}{\mathcal B}(B\to ZK)\times{\mathcal
B}(Z\to\psi'\pi)=(4.1\pm1.0\pm1.3)\times 10^{-5}, \ee
%&&&&&&&&&&&&&&&&&&&&&
the partial width $\Gamma(Z\to\psi'\pi)$ should also be quite
significant. Together with the "$J/\psi$-suppressed problem", this
implies that the mechanism, which induces these hidden-charm decay
modes, should have some nontrivial features. We will examine these
modes in the re-scattering
model~\cite{Cheng05_ReSC,Zhao05_ReSC,Liu07_X3872_ReSC,Meng07_X3872_ReSC},
in which the $D^*D_1^{(\prime)}$ component is re-scattered into
$\psi^{(\prime)}\pi$ by one $D^{(*)}$ meson exchange. In this
picture, the form factor suppression, which accounts for the
off-shell effects of the intermediate meson, will favor $\psi'$
because its mass is closer to the $D(*)\bar{D}(*)$ thresholds than
that of $J/\psi$. We find that the lager width of $Z\to\psi'\pi$ and
the $J/\psi$-suppressed problem may be explained if one choose
adequate parameters.
%On the other hand, we will discuss the
%production mechanism of the $Z$ resonance briefly.

Needless to say, as charged states, the $Z^{\pm}$ should be in a
isospin-triplet, and their isospin partner $Z^0$ should exist. In
our analysis, we will focus on $Z^+$. However, all the results can
be applied to $Z^0$ directly if the interaction between $Z$ and
$D^*D_1^{(\prime)}$ has isospin symmetry.

We give the model to describe the decays of $Z^+(4430)$ in Sec. II.
The numerical analysis and discussions are given in Sec. III. And a
summary is given in Sec. IV.

\section{The Decays of $Z^+(4430)$}

As a charged resonance, the $Z^+(4430)$ can be coupled to either
$\bar{D}^{*0}D_1^{(\prime)+}$ or $D^{*+}D_1^{(\prime)0}$. We will
describe their interactions by the following effective Lagrangians
with isospin symmetry:
%&&&&&&&&&&&&&&&&&&&&&
\bqa {\mathcal L}^0_{ZD_1^{(\!\prime\!)}D^*}\!\!\! &=&\!\!
g^0_{ZD_1^{(\!\prime\!)}D^*} Z (D^*\cdot D_1^{(\prime)\dagger})+h.c.~~~~~~~~\mbox{for}~0^-,\label{L^0_ZDD}\\
{\mathcal L}^1_{ZD_1^{(\!\prime\!)}D^*}\!\!\!  &=&\!\!
ig^1_{ZD_1^{(\!\prime\!)}D^*}\epsilon^{\mu\nu\rho\sigma}D^{*}_\mu
D^{(\prime)\dagger}_{1\nu} \partial_\rho Z_\sigma\!+\!
h.c.~~\!\!\mbox{for}~1^-,\label{L^1_ZDD} \eqa
%&&&&&&&&&&&&&&&&&&&&&
where the coupling constants $g^{0,1}_{ZD_1^{(\!\prime\!)}D^*}$ are
blind to the charge of $D$ meson, but there can be significant
difference between these for $D_1$ and for $D_1'$ due to the
dynamics in $D^*D_1^{(\prime)}$ systems.
%Furthermore, the values of
%these coupling constants are expected small than those between
%charmonia and $D$ mesons since the interaction between $D^*$ and
%$D_1(D_1^\prime)$ is general weak.
As we have mentioned in Sec.~I,
we only introduce S-wave coupling between $Z$ and
$D^*D_1^{(\prime)}$ with $J^P(Z)=0^-$ and $1^-$ in
Eq.~(\ref{L^0_ZDD}) and (\ref{L^1_ZDD}), respectively. Then, the
study on the decays of $Z$ resonance is exigent to clarify its
quantum number even its components.

In this section, we will focus on two primary decay modes of $Z$,
i.e., the open-charm mode $Z\to D^*D^*\pi$ and the hiden-charm mode
$Z\to J/\psi(\psi')\pi$. For the open-charm mode, we assume that one
of the two $D^*$s arise from the decay of $D_1(D_1^\prime)$
component and the propagation of virtual  $D_1(D_1^\prime)$ is
described by the Breit-Wigner propagator. For the hiden-charm decay
mode, we use re-scattering mechanism, which was used in the analysis
of $X(3872)\to J/\psi \rho$~\cite{Meng07_X3872_ReSC}.

\subsection{$Z^+\rightarrow \psi^{(\prime)}\pi^+$}

\begin{figure}[t]
\begin{center}
\vspace{-2.7cm}
 \hspace*{-3.6cm}
\scalebox{0.5}{\includegraphics[width=32cm,height=40cm]{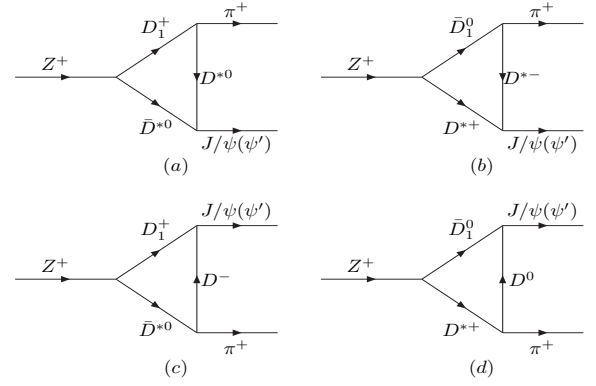}}
\end{center}
\vspace{-12.0cm}\caption{The decay diagrams for $Z^+(4430)\to
\bar{D}^{*0}D_1^{+}(D^{*+}D_1^{0})\to
J/\psi(\psi')\pi^+$.}\label{Fig:ZPsiPi}
\end{figure}

In the re-scattering mechanism, the decay $Z^+\to
J/\psi(\psi')\pi^+$ can arise from exchange of a $D^{(*)}$ meson
between $D^*$ and $D_1(D_1^\prime)$. We will not take into account
higher excited $D$ meson medium states because their contributions
will be suppressed by the form factors intensively. The Feynman
diagrams for $Z^+(4430)\to \bar{D}^{*0}D_1^{+}(D^{*+}D_1^{0})\to
J/\psi(\psi')\pi^+$ are shown in Fig.~\ref{Fig:ZPsiPi}. The diagrams
for $D^*D_1'$ re-scattering are the same.

To evaluate the amplitudes, we need the following
effective~\cite{Casalbuoni,Falk92-D1D2} Lagrangians:
%%%%%%%%%%%%%%%%%%%%%%%%%%%%
\begin{eqnarray}
\mathcal{L}_{\psi D^*D^*}&=&-i g_{\psi D^* D^*} \Bigl\{ \psi^\mu
\left(\partial_\mu D^{*\nu} {D}_\nu^{*\dagger} -D^{*\nu}\partial_\mu {D}_\nu^{*\dagger} \right)\nonumber\\
&&+ \psi^\nu D^{*\mu}\partial_\mu{D}^{*\dagger}_{\nu} -
\psi_\nu\partial_\mu
D^{*\nu}  {D}^{*\mu\dagger}  \mbox{}   \Bigr\},\label{L_psiD*D*}\\
\mathcal{L}_{\psi D_1^{(\!\prime\!)}D}&=& g_{\psi
D_1^{(\!\prime\!)}D} (\psi\cdot D_1^{(\prime)})D^\dagger
+h.c.,\label{L_psiD_1D}\\
\mathcal{L}_{D_1D^*\pi}&=& ig_{D_1D^*\pi}[-3D^*_\mu D_{1\nu}^\dagger \partial^\mu\partial^\nu\pi+(D^*\!\cdot\! D_1^\dagger)\partial^2\pi\nonumber\\
&&-\frac{1}{m_{D^*}m_{D_1}}\partial_\mu D^{*\rho}\partial_\nu D_{1\rho}^\dagger\partial^\mu\partial^\nu\pi]+h.c.,\label{L_D_1Dpi}\\
\mathcal{L}_{D_1'D^*\pi}&=& ig_{D_1'D^*\pi} D^{*\mu}\partial_\nu D_{1'\mu}^\dagger\partial^\nu\pi+h.c.,\label{L_D_1'Dpi}\\
\mathcal{L}_{D^*D\pi}&=& ig_{D^*D\pi}D^{*}_\mu\partial^\mu \pi
D^\dagger+h.c..\label{L_D*Dpi}
\end{eqnarray}
%%%%%%%%%%%%%%%%%%%%%%%%%%%%
Again, we assume that the isospin symmetry is preserved in
(\ref{L_psiD*D*})-(\ref{L_D*Dpi}). That is, the coupling constants
are blind to the flavor of light quarks and the coupling constants
in (\ref{L_D_1Dpi})-(\ref{L_D*Dpi}) should product by a factor of
$1/\sqrt2$ for interactions involving $\pi^0$.

For convenience, we define the following ratios:
%&&&&&&&&&&&&&&&&&&&&&
\be r_{D^*D^*}=\frac{g_{\psi' D^* D^*}}{g_{\psi D^*
D^*}},~~~~r_{D_1^{(\!\prime\!)}D}=\frac{g_{\psi
D_1^{(\!\prime\!)}D}}{g_{\psi
D_1^{(\!\prime\!)}D}}.\label{r:defination}\ee
%&&&&&&&&&&&&&&&&&&&&&&
In our analysis in Sec. III, we will not distinguish them seriously,
but call them as "$r$" totally.

All the coupling constants will be determined in the next section.
However, it is necessary to emphasize here that the determinations
will not account for the off-shell effect of the exchanged $D(D^*)$
meson, of which the virtuality can not be ignored. Such effects can
be accounted for by introducing, e.g., the
monopole~\cite{Cheng05_ReSC} or dipole~\cite{Zhao05_ReSC} form
factors for off-shell vertexes. Let $q$ denote the momentum
transferred and $m_i$ the mass of exchanged meson, the form factor
can be written as
%%%%%%%%%%%%%%%%%%%%%%
\begin{eqnarray}\label{formfactor1}
\mathcal{F}(m_{i},q^2)=\bigg(\frac{\Lambda^{2}-m_{i}^2
}{\Lambda^{2}-q^{2}}\bigg)^n,
\end{eqnarray}
%%%%%%%%%%%%%%%%%%%%%%
and the cutoff $\Lambda$ can be parameterized as
%%%%%%%%%%%%%%%%%%%%
\begin{eqnarray}\label{Lamvda}
\Lambda(m_{i})=m_{i}+\alpha \Lambda_{QCD}.
\end{eqnarray}
%%%%%%%%%%%%%%%%%%%%%
As we have mentioned, this form factor may play a important role in
understanding the "$J/\psi$-suppressed problem". So our numerical
results will be very sensitive to the values of $n$ and $\alpha$.
Since increasing of $n$ is equivalent to decreasing of $\alpha$ in a
large kinematical region, we will choose to fix $\alpha=3$, and to
vary the value of $n$ from 1 to 2.

We are now in a position to compute the diagrams in
Fig.~\ref{Fig:ZPsiPi}.  If the $Z^+(4430)$ lies above the
$D^*D_1(D^*D_1')$ threshold, in the processes $Z(p_X,\epsilon_X)\to
D_1(D_1^\prime)(p_{1},\epsilon_1)+D^{*}(p_{2},\epsilon_2)\to
\psi(\psi^\prime)(p_{3},\epsilon_3)+\pi(p_{4})$, where the momenta
$p$ and polarization vectors $\epsilon$ are denoted explicitly for
the mesons,  we can calculate the absorptive part (imaginary part)
of Fig.~\ref{Fig:ZPsiPi} and find it to be given by
%**************
\bqa\label{Abs:CutRule}
\textbf{Abs}_i&=&\frac{|\vec{p}_2|}{32\pi^2m_Z}\int d\Omega
\mathcal{A}_i(Z\to D^*\bar{D}_{1}(D_1^\prime))\nonumber\\
&&\times \mathcal{A}_i(D^*\bar{D}_{1}(D_1^\prime)\to
\psi(\psi^\prime)\pi), \eqa
%**************
where $i=(a,b,c,d)$ and $\vec{p}_2$ is the 3-momentum of the
on-shell $D^*$ meson in the rest frame of $Z(4430)$.

Neglecting the mass difference between charged and neutral $D$
mensons, the amplitudes of Fig.~\ref{Fig:ZPsiPi}(a) and
\ref{Fig:ZPsiPi}(c) will be equal to those of
Fig.~\ref{Fig:ZPsiPi}(b) and \ref{Fig:ZPsiPi}(d), respectively.
However, the absorptive part of the amplitude is indeed sensitive to
the difference between the threshold of
$\bar{D}^{*0}D_1^{(\prime)+}$ and of $D^{*+}D_1^{(\prime)0}$. This
is because the absorptive part in (\ref{Abs:CutRule}) is
proportional to the phase space factor $|\vec{p}_2|/m_2$ and is
sensitive to the exact position of the threshold. On the other hand,
the absorptive part is strongly suppressed by this phase space
factor. Even smeared with the Breit-Wigner distribution function
%&&&&&&&&&&&&&&&&&&&&&&&&7
\be \frac{1}{\pi}\frac{\sqrt
t\Gamma_i(t)}{(t-m_i^2)^2+t\Gamma_i(t)^2},~~~i=D_1,~D_1',\label{BWdistribution}\ee
%&&&&&&&&&&&&&&&&&&&&&&&&&&
the absorptive part is still less than the real part of the
amplitude and can be neglected.

The evaluation of the real part of the amplitude is difficult to be
achieved. We will follow Ref.~\cite{Meng07_X3872_ReSC} to obtain it
from absorptive part via the dispersion relation:
%&&&&&&&&&&&&&&&&&&&&&&7
\be \label{Dispertion relation}
\textbf{Dis}(m_Z^2)=\sum^d_{i=a}\frac{1}{\pi}\int_{th^2}^{\infty}\frac{\textbf{Abs}_i(s')}{s'-m_Z^2}ds'
,\ee
%&&&&&&&&&&&&&&&&&&&&&&&&&
where $th=m_{1}+m_2$ is the threshold. To proceed, we need a
reasonable cutoff for the integral in (\ref{Dispertion relation}).
We choose to add a form factor on the absorptive
part~\cite{Pennington-charmonium,zhang07_ReSC}:
%&&&&&&&&&&&&&&&&&&&&&
\be\label{formfactor2} \textbf{Abs}_i(s)\rightarrow
\textbf{Abs}_i(s)e^{-\beta|\vec{p}_2|^2},\ee
%&&&&&&&&&&&&&&&&&&&&&
where the cutoff $\beta$ can be related with the effective radius of
interaction $R$ by $\beta=R^2/6$. The authors of
Ref.~\cite{Pennington-charmonium} choose $\beta=0.4$ GeV$^{-2}$,
which corresponds to $R\thickapprox0.3$ fm. We will choose
$\beta=0.4\mbox{-}1.0$ GeV$^{-2}$ for our numerical analysis.

\subsection{$Z\rightarrow D^*D^*\pi$}

\begin{figure}[t]
\begin{center}
\vspace{-2.9cm}
 \hspace*{-1.3cm}
\scalebox{0.5}{\includegraphics[width=22cm,height=34cm]{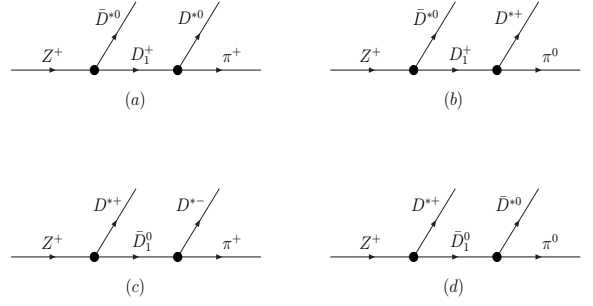}}
\end{center}
\vspace{-10.4cm}\caption{The diagrams for $Z^+(4430)\to
D^*D^*\pi^+$.}\label{Fig:ZDDpi}
\end{figure}

For the cascade decay $Z\to D^* D_1(D_1')\to D^*D^*\pi$, the Feynman
diagrams four $Z^+$ decay modes are illustrated in
Fig.~\ref{Fig:ZDDpi}. The four diagrams in Fig.~\ref{Fig:ZDDpi} are
related by the isospin symmetry, and one can easily obtain the
approximate relation:
%&&&&&&&&&&&&&&&&&&&&&&&&&
\bqa \Gamma(Z^+\to D^{*+}D^{*-}\pi^+)&\approx&\Gamma(Z^+\to
\bar{D}^{*0}D^{*0}\pi^+)\nonumber\\&\approx&\frac{1}{2}\Gamma(Z^+\to
\bar{D}^{*0}D^{*+}\pi^0).\nonumber\eqa
%&&&&&&&&&&&&&&&&&&&&&&&&&

Since in our model, one $D^*$ directly comes from the virtual
$D_1(D_1')$ component and have no interference with the other $D^*$,
we can use the formula for cascade decay to evaluate the partial
width as
%&&&&&&&&&&&&&&&&&&&&
\bqa\label{Cascade decay formula} \Gamma_{Z^+\to
D^{*+}D^{*-}\pi^+}&=&\frac{1}{\pi}\int^{(m_Z-m_{D^*})^2}_{(m_{\pi}+m_{D^*})^2}ds\sqrt
s\nonumber\\
&&\hspace*{-2.3cm}\times\frac{\Gamma_{Z^+\to
D^{*+}\bar{D}_1^{(\prime)0}\pi^+}(s)\Gamma_{\bar{D}_1^{(\prime)0}\to
D^{*-}\pi^+}(s)}{(s-m_{D_1^{(\prime)}}^2)^2+(\sqrt{s}\,\Gamma_{D_1^{(\prime)}})^2}.\eqa
%&&&&&&&&&&&&&&&&&&&&

For the neutral state $Z^0$, the number of the diagrams for decay
$Z^0\to D^*D^*\pi$ is eight. However, the total width of open-charm
decay of $Z^0$ should be approximation equal to that of $Z^+$ in our
model.

\section{Numerical Results and Discussions}

The coupling constants of $DD\pi$ in(\ref{L_D_1Dpi})-(\ref{L_D*Dpi})
can be related to the well-known parameters in Heavy Meson Chiral
Lagrangian through the relations~\cite{Casalbuoni,Falk92-D1D2}:
%&&&&&&&&&&&&&
\bqa\label{g_DDpi:value}
g_{D^*D\pi}&=& 2\frac{g}{f_\pi}\sqrt{m_{D^*}m_{D}},~~g_{D_1'D^*\pi}=2\frac{h}{f_\pi}\sqrt{m_{D^*}m_{D_1'}},\nonumber\\
&&g_{D_1D^*\pi}=\frac{2}{\sqrt6}\frac{h_1+h_2}{f_\pi\Lambda_\chi}\sqrt{m_{D^*}m_{D_1}},
\eqa
%&&&&&&&&&&&&&&&
where $f_\pi$ is the decay constant of $\pi$ and $\Lambda_\chi$ is
chiral symmetry breaking scale. The parameters in
(\ref{g_DDpi:value}) can be roughly estimated through the
measurement of the width of $D^{*+}(D_1',D_1)$~\cite{PDG06}. Using
the central value of these widthes, the parameters $|h_1+h_2|$, $h$
and $g$ can be evaluated and we list them in Tab.~\ref{parameters}.

The evaluations of the coupling constants $g_{\psi D^*D^*}$ and
$g_{\psi D_1^{(\prime)}D}$ are somewhat difficult since there are
hardly any experimental data to be used. The two things we can use
are symmetry and model. For $g_{\psi D^*D^*}$, we can relate it to
$g_{\psi DD}$ through heavy quark symmetry, and then estimate its
value with the help of Vector Meson Dominant (VMD)
model~\cite{Deandrea03}:
%&&&&&&&&&&&&&&&&
\be\label{g_psiD*D*:VMD} g_{\psi D^*D^*}\approx g_{\psi DD}\approx
\frac{m_{\psi}}{f_\psi}\approx8,\ee
%&&&&&&&&&&&&&&&&&&&
where $f_\psi$ is the decay constant of $\psi$. For $g_{\psi
D_1(D_1')D}$, we take it as large as $g_{\psi(4415) D_1D}$, which
can be estimated through the prediction of $\Gamma(\psi(4415)\to D_1
D)$ in $^3P_0$ model~\cite{Barnes}.

The coupling constant $g^{1}_{ZD_1D^*}$ and the re-scaled one
$g^{0}_{ZD_1^{\prime}D^*}/\sqrt{m_Zm_{D^*}}$ should be smaller than
$g_{\psi D^*D^*}$ since as a resonance in $D_1^{(\prime)}D^*$, $Z$
must couples with $D_1^{(\prime)}D^*$ through some weak dynamics
(say, $\pi$-exchange). So we choose
$$g^{1}_{ZD_1^{(\prime)}D^*}=1.5,~~~g^{0}_{ZD_1^{(\prime)}D^*}=5\mbox{~GeV}$$
in Tab.~\ref{parameters}. And these parameters give the prediction
on partial widths of open-charm decays of $Z(4430)$:
%&&&&&&&&&&&&&&&&&&&&&&&&&&&&&&&&&&&
\bqa\label{Th:ZtoDDpi}\Gamma(Z(0^-)\to D_1D^*\to
D^*D^*\pi)&=&25\mbox{~MeV},\nonumber\\
\Gamma(Z(0^-)\to D_1'D^*\to
D^*D^*\pi)&=&37\mbox{~MeV},\nonumber\\
\Gamma(Z(1^-)\to D_1D^*\to
D^*D^*\pi)&=&32\mbox{~MeV},\\
\Gamma(Z(1^-)\to D_1'D^*\to D^*D^*\pi)&=&46\mbox{~MeV}.\nonumber
 \eqa
 On the other hand, we also consider about that the resonance decays to
 $D_1^{(\prime)}D^*$ directly. After smeared with the Breit-Wigner
 distribution functions defined in (\ref{BWdistribution}), all the
 partial widthes is about 30 GeV. So, we will use the numbers in
 (\ref{Th:ZtoDDpi}) in our discussions.

%--------------
\begin{table}[t]
\caption{Parameters used in the calculations.}
%**************
\label{parameters}
\begin{tabular}{c|c|c|c|c}
\hline\hline $g_{\psi D^*D^*}$   &    $g_{\psi
D_1^{(\!\prime\!)}D^*}$ & $|h_1+h_2|/\Lambda_\chi$ & $h$ & $g$
\\ \hline
    8   &    5 GeV   &  1 Gev$^{-1}$ &   0.7  &   0.6
    \\\hline\hline
       $g^0_{Z D_1^{(\!\prime\!)}D}$   &  $g^1_{Z D_1^{(\!\prime\!)}D^*}$ &   $f_\pi$  &  $\Lambda_{QCD}$ &
$r$   \\\hline
   5 Gev   &   1.5  &   132 MeV  &  220 MeV  &  2 \\
\end{tabular}
%*******************
\end{table}
%--------------

As for the hidden-charm decay $Z^+\to \psi'\pi^+$, the width is
squarely dependent on the value of $r$, which is defined in
(\ref{r:defination}). It have been argued that the coupling between
$DD$ system and excited charmonium is generally not weak comparing
with that for $J/\psi$. For example, the experimental data tell us
that the coupling constant $g_{\psi(3770)DD}\approx24$, which is
about 3 times of $g_{\psi DD}$. That is, the value of $r$ tends to
be greater than 1. In our discussion, we choose relative large value
for $r$ in Tab.~\ref{parameters} to examine the possibilities to
understand the "$J/\psi$-suppression problem" in our model.

Evaluating the amplitudes in (\ref{Dispertion relation})
numerically, we find that the contributions from $D^*D_1'$ channel
are far less than those from $D^*D_1$ channel. This is mainly
because the large width of $D_1'$ reduces the threshold effects
consumedly. On the other hand, the contributions arising from
Fig.~1(c)-1(d) are less than those from Fig.~1(a)-1(b) by a factor
of 15-30. This is probably because the value coupling constant
$g_{\psi D_1D}$ which we choose in Tab.~\ref{parameters} is small.
We can see it through re-scaling $g_{\psi D_1D}$ by
$\sqrt{m_{D_1}m_{D}}$ and getting the number is 2.3, which is far
less than that of $g_{\psi D^*D^*}$.

%--------------
\begin{table}[t]
\caption{The predictions of $\Gamma(Z^+)\to\psi^{(\prime)}\pi^+$ and
their dependence on (n, $\beta$). Here, we only involve the
contributions arising from Fig.~1(a)-1(b) in $D^*D_1$ channel.}
%**************
\label{Tab:hiden-charm width:D_1}
\begin{tabular}{c|c|c|c|c|c|c}
\hline   (n, $\beta\times\mbox{GeV}^2$) & \multicolumn{2}{c|}{(1.0,
1.0)}& \multicolumn{2}{c|}{(1.5, 0.6)}& \multicolumn{2}{c}{(2.0,
0.4)}\\\hline
     $J^P(Z)$ &   ~$0^-$~   &  ~$1^-$~ &    ~$0^-$~   &  ~$1^-$~ &    ~$0^-$~   &  ~$1^-$~
    \\\hline
      $ \Gamma_{\psi\pi^+}$(MeV) &2.4&4.7&0.68&1.4&0.19&0.36 \\
       $\Gamma_{\psi'\pi^+}$(MeV) &1.6&6.0&1.1&3.8&0.64& 1.9\\
\end{tabular}
%*******************
\end{table}
%--------------

The numerical result of the partial widths
$\Gamma(Z^+)\to\psi^{(\prime)}\pi^+$ are listed in
Tab.~\ref{Tab:hiden-charm width:D_1}. Here, the contributions
involved are only that arising from Fig.~1(a)-1(b) in $D^*D_1$
channel. From Tab.~\ref{Tab:hiden-charm width:D_1}, one can find
that the prediction on $R_{\psi'/\psi}$ increases with the parameter
n increasing as one expects. For the $0^-$ candidate, the ratio
$R_{\psi'/\psi}\approx3$ at $n=2$, which have implied the "$J/\psi$
suppression" in some degree. However, the prediction on the partial
width at $n=2$ is too small comparing with the open-charm one in
(\ref{Th:ZtoDDpi}), which indicates that: $\mathcal{B}(Z^+(0^-)\to
\psi'\pi^+)\sim2.5\%$ and $\mathcal{B}(B\to
ZK)\sim1.6\times10^{-3}\%$. So the experimental data seems to
disfavor the $0^-$ candidate.

For the $1^-$ candidate, when $n=2$, the prediction on the ratio
$R_{\psi'/\psi}\approx5$ with $\mathcal{B}(Z^+(0^-)\to
\psi'\pi^+)\sim6\%$, which are better than that of $0^-$ candidate
and are roughly consistent with experimental data.

If we choose $\alpha<3$ ~\cite{zhang07_ReSC} and fix $n=2$, it is
easy to get a large value prediction on $R_{\psi'/\psi}$ which can
be consistent with the experimental constraint in
(\ref{Ex:R_psi'/psi}). As a price to pay, the prediction of
$\Gamma_{\psi'\pi^+}$ will be decreased. On the other hand, if the
coupling constant $g_{\psi D_1D}$ is not as small as that given in
Tab.~\ref{parameters}, or if we can evaluate it by re-scaling
$g_{\psi D^{(*)}D^{(*)}}\sqrt{m_{D_1}m_D}$, then the contributions
arising from Fig.~1(c)-1(d) will be comparable to those from
Fig.~1(a)-1(b), and the width $\Gamma_{\psi'\pi^+}$ might be
enhanced significantly since the interference effects turn to be
important. Moreover, in Fig.~1(c)-1(d), the couplings of
$\psi^{(\prime)}DD$ are S-wave while those in Fig.~1(a)-1(b) are
P-wave. So, the configurations in Fig.~1(c)-1(d) are more in favor
of $\psi'$-productions than those in Fig.~1(a)-1(b).

In conclusion, our calculations imply that the $Z^+(4430)$ state is
likely to be a resonance in S-wave $D^*D_1$ scattering, and the
spin-parity favors $1^-$ but can not rule out $0^-$. Further studies
on its production mechanism are needed.

\section{Summary}

In summary, we study both the open-charm and the hidden-charm decays
of $Z(4430)$ based on the assumptions that it is a
$D_1^{(\prime)}D^*$ resonance in S-wave. The partial width of
open-charm decay is dominant and can be adjusted to be 30-40 MeV.
Then, we find that the S-wave threshold effects are significant in
$D_1D^*$ channel and are suppressed intensively in $D_1'D^*$ channel
due to the larger width of $D_1'$. For the $1^-$ candidate, choosing
parameters aptly, we can get $\Gamma(Z^+\to
\psi'\pi^+)\thickapprox2$ MeV with $R_{\psi'/\psi}\thickapprox5.3$,
which could be roughly consistent with experimental data. But the
$0^-$ candidate can not be ruled out by our calculations. Further
studies on the production mechanism of $Z(4430)$ are needed.

$Note.$ When this manuscript was written, two papers about the
$Z^+(4430)$ appeared.  In \cite{Rosner07:Z4430}, a similar idea to
ours was proposed; while in \cite{4q} a tetraquark explanation was
suggested.

\begin{acknowledgments}
We wish to thank X. Liu, X.Y. Shen, C.Z. Yuan, H.Q. Zheng, and S.L.
Zhu for helpful discussions. This work was supported in part by the
National Natural Science Foundation of China (No 10421503, No
10675003), the Key Grant Project of Chinese Ministry of Education
(No 305001), and the Research Found for Doctorial Program of Higher
Education of China.
\end{acknowledgments}
%\newpage %Just because of unusual number of tables stacked at end

\bibliography{apssamp}% Produces the bibliography via BibTeX.

\end{document}